\documentclass[10pt, conference]{IEEEtran}
\IEEEoverridecommandlockouts

\usepackage{cite}
\usepackage[most]{tcolorbox}
\usepackage{amsmath,amssymb,amsfonts}
\usepackage{algorithmic}
\usepackage{graphicx}
\usepackage{textcomp}
\usepackage{xcolor}
\usepackage{booktabs}
\usepackage{hyperref}
\usepackage{siunitx}
\usepackage{multirow}

\def\BibTeX{{\rm B\kern-.05em{\sc i\kern-.025em b}\kern-.08em
    T\kern-.1667em\lower.7ex\hbox{E}\kern-.125emX}}
\begin{document}

\title{RM -RF: Reward Model for Run-Free Unit Test Evaluation}

\author{

\IEEEauthorblockN{
Elena Bruches\IEEEauthorrefmark{1},
Daniil Grebenkin\IEEEauthorrefmark{1},
Mikhail Klementev\IEEEauthorrefmark{1},
Vadim Alperovich\IEEEauthorrefmark{2},
Roman Derunets\IEEEauthorrefmark{1},
Dari Baturova\IEEEauthorrefmark{1}
}

\IEEEauthorblockN{
Georgy Mkrtchyan\IEEEauthorrefmark{2},
Oleg Sedukhin\IEEEauthorrefmark{1},
Ivan Bondarenko\IEEEauthorrefmark{3},
Nikolay Bushkov\IEEEauthorrefmark{2},
Stanislav Moiseev\IEEEauthorrefmark{2}
}

\IEEEauthorblockA{
\IEEEauthorrefmark{1}\textit{Siberian Neuronets LLC}, Novosibirsk, Russia\\
\IEEEauthorrefmark{2}\textit{T-Technologies}, Moscow, Russia\\
\IEEEauthorrefmark{3}\textit{Novosibirsk State University}, Novosibirsk, Russia\\
\{bruches, grebenkin, klementev, derunets, baturova, sedukhin\}@sibnn.ai\\
\{v.alperovich, g.p.mkrtchyan, n.bushkov, s.moiseev\}@t-tech.dev\\
i.bondarenko@g.nsu.ru
}
}

\maketitle

\begin{abstract}

We present \texttt{\textbf{RM -RF}}, a lightweight reward model for run-free evaluation of automatically generated unit tests. Instead of repeatedly compiling and executing candidate tests, \texttt{RM -RF} predicts — from source and test code alone — three execution-derived signals: (1) whether the augmented test suite compiles and runs successfully, (2) whether the generated test cases increase code coverage, and (3) whether the generated test cases improve the mutation kill rate. To train and evaluate \texttt{RM -RF} we assemble a multilingual dataset (Java, Python, Go) of focal files, test files, and candidate test additions labeled by an execution-based pipeline, and we release an associated dataset and methodology for comparative evaluation. We tested multiple model families and tuning regimes (zero-shot, full fine-tuning, and PEFT via LoRA), achieving an average F1 of 0.69 across the three targets. Compared to conventional compile-and-run instruments, \texttt{RM -RF} provides substantially lower latency and infrastructure cost while delivering competitive predictive fidelity, enabling fast, scalable feedback for large-scale test generation and RL-based code optimization.

\end{abstract}


\section{Introduction}

Automatic code generation is the autonomous creation of software artifacts using predefined algorithms or intelligent systems. Historically, automated code generation heavily relied on rule-based engines and templating mechanisms \cite{10.1145/1321631.1321646, 10.5555/2819009.2819139}, limiting their application to simple, repetitive tasks requiring precise specification of logic by developers.

However, over the past decade there has been an exponential increase in the capability of automatic code generation driven by advancements in large-scale language models (LLMs) integrated with deep learning and natural language processing technologies. State-of-the-art models, such as GPT \cite{openai2024gpt4technicalreport}, DeepSeek \cite{deepseekai2024deepseekv3technicalreport}, Qwen \cite{yang2025qwen3technicalreport} and others, demonstrate exceptional proficiency in interpreting natural-language instructions and synthesizing executable code that resolves complex programming challenges, leading to gains in programmer productivity.

Despite these breakthroughs, ensuring alignment between generated code and diverse user requirements remains a critical issue. Sophisticated methodologies, particularly that employ reinforcement learning (RL), integrate direct human feedback into the training loop, enabling fine-tuned outputs that conform more closely to practical usage contexts \cite{10.5555/3600270.3602281}. Nonetheless, gathering high-quality feedback incurs substantial costs in terms of effort and resources.

Another promising feature of RL-based approaches is their ability to exploit feedback directly from compilation \cite{bi-etal-2024-iterative} or unit testing \cite{liu2023rltfreinforcementlearningunit} stages during runtime. This mitigates dependency on manual oversight, simultaneously improving precision and scalability across various code generation tasks.

Finally, traditional validation measures, such as unit tests that compare produced outputs with expected results, provide robust means of evaluating correctness. However, executing build pipelines, running unit tests, and computing relevant metrics impose significant computational overhead, delaying the overall training and inference processes.

In light of these challenges, our study investigates an alternative method to reduce latency and scale up the adoption of automatic code generation technologies, contributing valuable insights toward advancing this field further.

We propose \textbf{RM -RF}, a simple and scalable approach to guide the unit test generation that evaluates the impact of each generated test case. The approach applies the pre-trained model that removes the need to compile and run tests each time during project testing stage. This model is trained to predict whether the particular test case makes the whole test suite correct (1) (i.e. it compiles and runs successfully), improves the code coverage (2) and mutation kill rate (3) or not.

To train this model, we prepared the dataset, which includes samples for three programming languages (Java, Python, and Go), both human-written and generated. To obtain the target values, all samples were run through specialized tools for the estimation of code coverage and mutation kill rate. Each sample includes the focal file and test file contents as a context and a new added test case, which is intended to be evaluated.

We conducted a series of experiments, using different family and size models such as Qwen and Codestral. Moreover, we experienced with different types of targets. We compared zero-shot, full-parameter fine-tuning, and PEFT (parameter-efficient fine-tuning) with LoRA \cite{hu2021loralowrankadaptationlarge}, achieving the 0.69 average F1-score on three targets.  

Providing the necessary feedback in the form of a set of metrics, one can assess whether the test case is useful or not while training the model or generating the code snippet.

The key contributions of this work are as follows:
\begin{enumerate}
    \item \textbf{Curated Real-world Dataset:} We curate and prepare a comprehensive dataset specifically designed for predicting (1) code correctness, (2) code coverage ratios, and (3) mutation kill rates. This dataset comprises well-defined training and validation splits. Additionally, we create a dedicated holdout set containing test cases generated by state-of-the-art LLMs, ensuring robust evaluation under realistic application conditions.
    \item \textbf{RM -RF:} We introduce a lightweight yet powerful reward model engineered to assess the utility of generated test cases efficiently. Designed to support Java, Python, and Go simultaneously, \textbf{RM -RF} eliminates the need for project builds and runtime executions, thereby reducing latency and resource costs significantly.
    \item \textbf{Empirical Analysis of SOTA LLMs:} We evaluate advanced LLMs in both zero-shot and fine-tuning modes to get a deeper insight how well they may process such task.
\end{enumerate}

The remainder of this paper is organized as follows. Section \ref{sec:back} reviews relevant works. Section \ref{sec:data-collection} presents our dataset design. Section \ref{sec:exp} describes the experiments with \textbf{RM -RF model}. Section \ref{sec:eval} reports our experimental results. Lastly, we discuss limitations and highlight directions for future work.

\section{Background}
\label{sec:back}

\subsection{Automated Test Generation}
Early automated unit test tools aimed at maximizing code coverage, such as Randoop \cite{10.1145/1297846.1297902} and EvoSuite \cite{10.1145/2025113.2025179}, laid the foundation for modern techniques. With the advent of large language models (LLMs), researchers began treating unit test generation as a generative task. For example, \cite{tufano2021unittestcasegeneration} used a sequence-to-sequence Transformer-based model (pretrained on code and text) to generate tests for Java. More recent work employs prompting and chain-of-thought strategies: ChatTester \cite{10.1145/3660783} leverages ChatGPT iteratively to refine self-generated tests, Libro \cite{10.1109/ICSE48619.2023.00194} applies post-processing steps to filter and rank LLM-generated tests. These approaches enable LLMs to generate or refine tests, but they often rely on heuristic prompting and struggle to consistently improve upon initial outputs.

\subsection{Evaluation of the Test Quality}
Recent studies and benchmarking efforts focus on evaluating LLMs’ ability to produce meaningful and executable tests, moving toward practical test quality metrics. For example, TestBench \cite{zhang2024testbench} evaluates LLMs across five key dimensions: syntactic correctness, compilation success, test validity, code coverage, and defect-detection rates. TestEval \cite{wang-etal-2025-testeval} examines three distinct dimensions: overall code coverage (cov@k), targeted line and branch coverage, and path coverage. These toolkits are usually targeting individual functions, have no support for assessing complete test suites at the file or repository level, and need time-consuming manual annotation or resorting to static ground-truth test oracles. 

Although simple to compute, code coverage alone does not reliably indicate a test suite’s ability to find real bugs. Mutation testing \cite{10.1016/S0164-1212(96)00154-9}, by contrast, injects artificial mutants (bugs) and measures the percentage caught (killed) by the tests. A higher mutation score typically reflects stronger tests. 
For example, MuTAP \cite{DAKHEL2024107468} shows that prompt augmentation, with surviving mutants improves test effectiveness, and  MutGen \cite{wang2025mutationguidedunittestgeneration} incorporates mutation feedback into its prompts to generate tests with high fault detection. 
A recent large-scale industrial study, Mutation-Guided LLM-based Test Generation at Meta (ACH) \cite{10.1145/3696630.3728544}, reinforces the practical value of mutation-based feedback. ACH integrates mutant generation, equivalent mutant detection, and LLM-based test synthesis, achieving over 70\% engineer acceptance rate across thousands of Android classes. Notably, many accepted tests did not increase line coverage yet substantially improved mutation scores — highlighting that coverage alone is an unreliable measure of test adequacy.

In addition, several benchmarks emulate real-world scenarios. DevBench ~\cite{li2024promptinglargelanguagemodels} simulates the entire software development lifecycle (SDLC), integrating tasks such as software design, environment setup, implementation, acceptance testing, and unit testing. 
Similarly, TestGenEval ~\cite{jain2025testgenevalrealworldunit} prioritizes the generation of comprehensive test suites, stressing initial test authorship, suite expansion, and coverage improvement. 
Our solution addresses the complete test maintenance cycle --- \emph{generation}, \emph{repair}, and \emph{update} of test suites. Unlike prior work focused on individual functions or test cases, it operates at the granularity of entire test files, reflecting realistic developer workflows.

\subsection{Reinforcement Learning and Reward Modeling in Code Generation}
Traditional outcome-based RL gives sparse rewards only when a complete program passes tests. To provide denser feedback, recent work on process-supervised reinforcement learning for code generation \cite{ye2025processsupervisedreinforcementlearningcode} automatically labels each line of generated code by compiling mutated versions, training a process-based reward model (PRM) that scores the “thought process” of code generation. Integrating this PRM into the RL loop (in PRLCoder) significantly outperforms outcome-only approaches, especially on complex tasks. Similarly, another recent work on process supervision–guided policy optimization \cite{dai2025processsupervisionguidedpolicyoptimization} trains a line-level PRM that imitates human debugging by delivering immediate feedback on partial code, leveraging it as a dense reward signal during training.

Moreover, StepCoder \cite{dou-etal-2024-stepcoder} introduces an RL framework that decomposes long programs into a curriculum of code completion subtasks, applying fine-grained optimization by masking unexecuted segments. Using compiler feedback and verified datasets to guide learning, StepCoder achieves stronger exploration and competitive benchmark results.
CodePRM \cite{li-etal-2025-codeprm} leverages execution results to label reasoning traces, collecting step-by-step thought–code pairs with pass/fail outcomes and training a model to score each reasoning step. Within a generate–verify–refine pipeline, it identifies reasoning errors and corrects them, yielding substantial gains over baseline methods.
$\mu$Code \cite{jain2025multiturn}, building on the Markov Decision Process (MDP) formulation \cite{1950bellman_mdp}, treats code generation as a one-step recoverable MDP: a generator proposes code, a verifier evaluates it via execution feedback, and the two alternate in a closed loop. This single-step reward scheme significantly outperforms prior hierarchical RL approaches.
Outcome Refining Process Supervision (ORPS) \cite{yu2025reasoning} unifies process and outcome rewards via tree search: it generates alternative code paths, profiles their execution, and performs self-critique, improving correctness by 26.9\% and enhancing code generation efficiency.
These RL-based techniques employ compiler and execution feedback — often through tests — to shape code generation, producing more accurate and robust programs.

Lastly, Regression Language Models for Code \cite{akhauri2025regressionlanguagemodelscode} introduces a model that predicts quantitative code metrics — such as memory footprint, GPU kernel latency, and model accuracy — directly from source text without execution. This approach aligns with our goal of estimating execution-derived test signals in a \emph{run-free manner}, demonstrating that text-based inference of code behavior can be both accurate and scalable.

\subsection{Learning and Testing via Adversarial RL}
Reinforcement learning has also been applied directly to unit test generation. For example, UTRL \cite{lee2025learninggenerateunittest}, an adversarial framework where two LLMs co-train: a test-generator LLM learns to produce tests that catch the code-generator LLM’s bugs, and the code-generator learns to satisfy those tests. In experiments, a Qwen-3B model trained with UTRL generated tests of higher quality than a supervised fine-tuned model, even outperforming GPT-4.1 on test generation. 
UTGen \cite{prasad2025learning} tackles the debugging scenario: authors train a model to create unit tests (inputs and expected outputs) that reveal flaws in faulty code, and UTDebug to iteratively use those tests for debugging. UTGen surpasses other LLM baselines in producing error-revealing tests, and incorporating UTGen’s feedback improves Qwen-32B’s pass rates on several benchmarks and demonstrates that UTGen is a better judge of code correctness.

In parallel, researchers have begun co-evolving the coder and tester. CURE \cite{wang2025coevolvingllmcoderunit} framework trains LLM-based code generator and LLM-based unit tester. together under RL: the coder writes solutions, the tester writes unit tests, and each learns from the other’s feedback without ground-truth code. Their ReasonFlux-Coder models show notable gains over similarly sized models like DeepSeek-Coder and Qwen-Coder, and the tester naturally transfers to improved test-time scaling.

 LLMs are increasingly being taught to generate, critique, and improve both programs and tests, using reinforcement signals from execution. Combining iterative critique (e.g. RefineCoder’s \cite{zhou2025refinecoderiterativeimprovinglarge} LLM-as-critic strategy and CTRL \cite{xie2025teaching} models trained to judge code outputs) with RL and co-evolution holds promise for more reliable code and test generation in the future.

To bridge the gap between informative execution signals and practical scalability, we introduce \textbf{RM -RF} approach, a lightweight pre-trained reward model that predicts — without compiling or running — whether a candidate unit test will allow the suite to compile and run, increase code coverage, and improve mutation kill rate. \textbf{RM -RF}  is trained on a multilingual dataset (Java, Python, Go) whose labels were obtained by executing each sample through our evaluation pipeline; we train and compare multiple model sizes and tuning regimes.

\section{Dataset Construction}
\label{sec:data-collection}
This section describes the motivation for choosing the target features for the Holdout dataset and training dataset, the process of unit test sample data collection, and splitting into subsets according to the target values. 

\subsection{Data evaluation}
\label{subsubsec:data-evaluation}

The set of data target features was chosen to show the unit test quality in the terms of its performance, focal file logic coverage and mutation robustness, and further filtration. These features of the tests can be measured by their correctness, coverage delta value, and mutation coverage delta value.

\textbf{Test correctness.} The positive label of test correctness considers lack of any type of runtime or syntax errors during unit test performance. It was estimated by the extraction of test running logs and their analysis. 

\textbf{Line coverage delta value} (\texttt{\(\Delta\)TestCov}). The line-level coverage is defined as follows:

\begin{equation}
     \texttt{TestCov} = \frac{\text{\# lines executed by tests}}{\text{\# total executable lines in the focal file}} \times 100,
\end{equation}

while \(\Delta\)TestCov is defined as the difference in line-level coverage between the initial coverage ratio of existing test cases and the final coverage ratio after adding the new test to the existing test suite:

\begin{equation}
    \Delta\texttt{TestCov} = \texttt{TestCov}_{\text{final}} - \texttt{TestCov}_{\text{initial}}
\end{equation}
The calculation of line coverage values depended on the project programming language: we used \texttt{coverage.py}\footnote{\url{https://github.com/nedbat/coveragepy}} for Python,  \texttt{JaCoCo}\footnote{\url{https://github.com/jacoco/jacoco}} plugin for Java and \texttt{cover} \footnote{\url{https://pkg.go.dev/cmd/cover}}  package for Go tests.

\textbf{Mutation coverage delta value} (\texttt{\(\Delta\)MutCov}). The aim of mutation analysis is to evaluate a test suite by introducing small, systematic changes (mutations) into the program under test and checking whether the tests detect the injected faults. Therefore, the mutation coverage delta value or "mutation kill rate" is estimated as a proportion of detectable (or failed) mutants that are killed by the test suite:

\begin{equation}
    \texttt{MutCov} = \frac{\text{\# failed mutants}}{\text{\# total mutants}} \times 100,
\end{equation}

Consequently, the mutation coverage delta value is estimated in a similar way to the line coverage delta value:

\begin{equation}
    \Delta\texttt{MutCov} = \texttt{MutCov}_{\text{final}} - \texttt{MutCov}_{\text{initial}}
\end{equation}
There are a few set of mutation analysis tools for every language. In our experiments it was performed with \texttt{mutpy}\footnote{\url{https://github.com/mutpy/mutpy}} library for Python, \texttt{PIT}\footnote{\url{https://pitest.org}} mutation testing system for Java and \texttt{go-mutesting}\footnote{\url{https://github.com/avito-tech/go-mutesting}} module for Go projects.

According to these set of metrics, every correct test that has a mutation coverage delta value or a line coverage delta value greater than zero is considered as well-written and useful.

\subsection{Data collection} 
The data collection pipeline for Holdout dataset and training dataset included several steps: GitHub project selection and downloading, project execution-based filtering and files' content-based filtering.

\subsubsection{Project selection}
We sourced open-source GitHub repositories using the GitHub API, applying a set of criteria to ensure that:

\begin{itemize}
    \item The repository contains test files with at least five focal-test file pairs;
    \item Data has not yet been included in LLM training sets (at least for the time being). The repositories for Holdout dataset had to show updates after January 1, 2020, and the specific source files needed their latest commit to be after January 1, 2025 (or January 1, 2024 for Java). The training dataset projects had to be updated at least after January 1, 2023.
    \item Data are available for usage according to a type of provided license, like permissive MIT and Apache-2.0;
    \item There is a confirmed usage experience of this projects, so for Holdout dataset we only considered repositories with a minimum of 40 stars (for training dataset $>$5 stars respectively) and feature contributions from at least two developers.
\end{itemize}

\subsubsection{Project execution-based filtering}

The training convergence and evaluation of reward model depend on the lack of dependencies which are not connected with the project setup. Thus, we excluded projects that necessitate additional dependencies, installation instructions, or the downloading of extra files for execution, ensuring the dataset remains scalable and reproducible. The unit tests were also executed twice to estimate performance stability and measure the runtime. The tests that run more then 30 seconds were excluded from the set to guarantee reasonable inference time during evaluation stage.

\subsubsection{Content-based filtering}
The last filtering stage included several steps to make samples suitable for training and evaluation:
\begin{itemize}
    \item The samples with fewer than two test cases were excluded;
    \item Removed pairs without at least one function of five+ executable lines;
    \item Discarded pairs with focal/test files $<$20 lines or above the 99th percentile (ignoring comments/blank lines);
    \item Removed automatically generated files to keep only human-authored samples;
    \item Excluded samples with $>$70\% comments in the focal file to focus on executable content.
\end{itemize}

Finally, we evaluated a suite of large language models (LLMs) on Holdout dataset, which presents the real-world data. The models tested included \texttt{DeepSeek-3.1}\cite{deepseekai2024deepseekv3technicalreport}, \texttt{Gemini-2.5-flash}\cite{2025arXiv250706261C}, \texttt{Devstral-small}\cite{mistral2025devstral}, \texttt{GPT-5}\cite{openai2025gpt41}, \texttt{GPT-OSS-120B}\cite{openai2025gptoss120bgptoss20bmodel}, and \texttt{Qwen3-Coder}\cite{yang2025qwen3technicalreport}. The tests generated by these models were then assessed using our evaluation pipeline to derive the relevant target metrics.

\subsection{Training dataset structure}

\begin{figure}[htbp]
\includegraphics[width=\columnwidth]{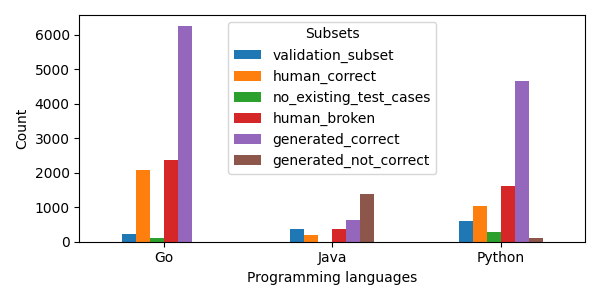}
\caption{Distribution of training dataset samples across programming languages}
\label{fig:langs_stats}
\end{figure}

\label{sec:dataset_structure}
The structure of the dataset used for training and validation is designed to improve the robustness of a reward model to different types of unit tests. The final set consists of 22285 samples and includes five training subsets and a single validation one, the samples were split into training and validation subsets with no repository overlap. The distribution of dataset samples across programming languages is presented on Figure \ref{fig:langs_stats}.

The samples were divided into several groups, based on test correctness labels and the origin of the unit test: there are human-written tests (with the \textit{human-} prefix) and LLM-written tests (with the \textit{generated-} prefix). The human-written tests include two sets:

\begin{itemize}
    \item \textbf{\textit{human\_correct}} (3303 samples, 14.82\% of total dataset). The samples from collected data which have successfully and syntactically correct existing tests and new tests.
    \item \textbf{\textit{human\_broken}} (4349 samples, 19.51\% of total dataset). The samples in this subset emulate a spectrum of potential errors commonly found in software code, showcasing how different LLMs address various types of test-related bugs. These test errors were generated using different LLMs, intentionally designed to disrupt tests based on specific input prompts. This set of samples includes instances of syntax errors, diverse runtime errors (e.g., invalid calls, missing dependencies), tests with redundant logic, tests lacking verification logic, and other scenarios.
\end{itemize}

The \textit{generated} part of the dataset was prepared with use of model \texttt{Qwen2.5-Coder-32B-Instruct} \cite{hui2024qwen2}. The goal of the generation prompt for this model is to produce additional unit tests for the specified focal file, building on the existing test code. This involves ensuring coverage of all key functionalities, exercising edge cases, adhering to testing best practices, and using fixtures and parameterization where appropriate. Processing of generation results considers identifying all test cases in each test file. Then, for every generated test case, we construct an isolated code variant by removing all other generated test cases. This procedure yields a dataset comprising multiple versions per test file, with each version containing exactly one test case. All of these samples were finally evaluated and divided to groups, based on test correctness metrics.

Therefore, the following subsets are created:

\begin{itemize}   
    \item \textbf{\textit{generated\_correct}} (11543 samples, 51.79\% of total dataset). These samples have successfully passed and syntactically correct generated tests.
    \item \textbf{\textit{generated\_not\_correct}} (1486 samples, 6.66\% of total dataset). Unlike the previous group, these tests have various runtime or syntax errors, which occur during evaluation.
\end{itemize}

The last group - \textbf{\textit{no\_existing\_test\_cases}} (396 samples, 1.77\%) - is created from the \textit{human} samples and \textit{generated}, which do not have the existing test cases filled. This subset aims to prevent the LLM's hallucination in the case when there is only a single new test. 

The \textbf{\textit{validation\_subset}} has got 1192 samples (5.34\% of total dataset).

\section{Experiments} 
\label{sec:exp}

\subsection{Training configuration}

We employ the Swift framework \cite{zhao2024swiftascalablelightweightinfrastructure} for fine-tuning and the vLLM library \cite{kwon2023efficient} for efficient inference.
For smaller models (7B parameters), we perform full supervised fine-tuning (SFT), while for larger models (20B parameters) we apply low-rank adaptation (LoRA) \cite{hu2022lora} to enable parameter-efficient fine-tuning.

We utilized NVIDIA A100 GPUs for all experiments. Hyperparameters were selected empirically based on preliminary runs.
SFT was conducted for 2 epochs with a learning rate of \num{1e-5}.
LoRA fine-tuning was performed for 3 epochs with a learning rate of \num{3e-5}, an adapter rank of 64, and an alpha coefficient of 16. Training experiments were carried out using the Brain Floating Point (BF16) half-precision format \cite{burgess2019bfloat16} and the Adafactor optimizer \cite{pmlr-v80-shazeer18a}.

\subsection{Model training prompt}

\begin{figure}
\begin{tcolorbox}[colback=gray!5, colframe=black!60, sharp corners=southwest, title=Reward Model Unit Test Evaluation Instruction]
\ttfamily
\small
\vspace{1em}
You are a code assistant that accepts a \{lang\} source file, test file and a new test case.
Your goal is to validate that the new test case is valid, and that it will increase the code coverage of the source file if it will be added to the existing test suite.

\vspace{1em}

Source file: 
\vspace{1em}
\{source\_file\_content\}

\vspace{1em}

Test file:
\vspace{1em}
\{test\_file\_content\}

\vspace{1em}

New test case:
\vspace{1em}
\{new\_test\_code\}

The output must be a YAML object equivalent to type TestValidation, according to the following Pydantic definitions:

\vspace{1em}

class TestValidation(BaseModel):

    is\_new\_test\_correct: bool 
    
    increase\_coverage: bool 

    increase\_mutation\_kill\_rate: bool

\end{tcolorbox}
\caption{Reward model unit test evaluation instruction (simplified version of prompt for binary targets)}
\label{fig:train_prompt}
\end{figure}

The input prompt (Figure \ref{fig:train_prompt}) for model's training consists of the focal file code, existing test cases file, and the new test file presented in diff-style format. This version of the unit test adopts a format inspired by tools such as \texttt{aider}\footnote{\url{https://github.com/Aider-AI/aider}} and other LLM-assisted code editing frameworks.

\subsection{Training targets}

The training targets for the reward model are defined at the end of each input prompt. We hypothesize that the choice of target formulation can directly influence the model’s performance. Accordingly, we examined multiple target formulations and constructed the corresponding versions of the dataset:

\begin{itemize}
    \item \textbf{Binary targets}: \texttt{correctness} (whether the test compiles and executes successfully), \texttt{coverage increase} (whether the test changes covers a larger portion of the source code logic), and \texttt{mutation increase} (whether the modification improves the mutation score and detects more potential bugs);
    \item \textbf{Float targets}: same for the \texttt{correctness}, while coverage and mutation correspond to the \texttt{\(\Delta\)TestCov} and \texttt{\(\Delta\)MutCov} targets respectively, as defined in Section~\ref{subsubsec:data-evaluation}.
\end{itemize}

On the initial stage, we also conducted experiments with \textbf{reversed binary targets}, where the model predicts \texttt{true} for coverage and mutation when the corresponding metrics decrease. However, in the final experiments, we used only the standard binary targets, as their behavior was similar to the reversed variant.

It is important to note that the targets appear to be interrelated (for instance, the values of TestCov and MutCov heavily rely on the correctness of tests). To validate this observation, we computed the correlations among these targets and discovered that they exhibit weak positive correlations (0.21 between Correctness and MutCov, 0.19 between Correctness and TestCov, and 0.35 between TestCov and MutCov). Among all programming languages analyzed, the correlation between TestCov and MutCov proves to be the most substantial. Additionally, an interesting finding is that the strength of target correlations is highest for the Go language, whereas it diminishes across other languages.

\begin{table*}[htbp]
  \centering
  \small
  \caption{Results on Validation subset} 
  \label{tab:main_results}        
\begin{tabular}{lcccc}
    \toprule                               
    \textbf{Model}                        &
    \textbf{Correctness}             &
    \textbf{TestCov}                     &
    \textbf{MutCov}                       &
    \textbf{Average}                      \\ 
    \midrule                              
    \multicolumn{5}{c}{\textbf{Binary targets}} \\
    \midrule
    \textit{Zero-shot}                    \\
    
    Qwen2.5-1.5B-Instruct & 0.52 & 0.54 & 0.53 & 0.53 \\
    Qwen2.5-Coder-7B-Instruct & 0.62 & 0.60 & 0.56 & 0.59 \\
    Qwen2.5-Coder-14B-Instruct & 0.64 & 0.62 & 0.61 & 0.62 \\
    
    Codestral-22B-v0.1 & 0.59 & 0.49 & 0.45 & 0.51 \\
    \midrule                              
    \textit{Fine-tuning}\\
    Qwen2.5-1.5B-Instruct & \underline{0.68} (+0.16) & 0.68 (+0.14) & 0.61 (+0.08) & 0.65 (+0.12) \\
    Qwen2.5-Coder-7B-Instruct & \textbf{0.69} (+0.07) & \textbf{0.76} (+0.16) & 0.63 (+0.07) & \textbf{0.69} (+0.10) \\
    Qwen2.5-Coder-14B-Instruct (LoRa) & 0.65 (+0.01) & 0.68 (+0.06) & 0.55 (-0.06) & 0.63 (+0.01) \\
    Codestral-22B-v0.1 (LoRa) & \underline{0.68} (+0.09) & 0.66 (+0.17) & 0.57 (+0.12) & 0.63 (+0.12)\\
    \midrule 
        \multicolumn{5}{c}{\textbf{Float targets}} \\
    \midrule
    \textit{Zero-shot} \\
    Qwen2.5-1.5B-Instruct & 0.52 & 0.51 & 0.52 & 0.52 \\
    Qwen2.5-Coder-7B-Instruct & 0.62 & 0.53 & 0.48 & 0.54 \\
    Qwen2.5-Coder-14B-Instruct & 0.62 & 0.59 & 0.49& 0.56 \\
    Codestral-22B-v0.1 & 0.56 & 0.51 & 0.51 & 0.52 \\
    \midrule
    \textit{Fine-tuning} \\
    Qwen2.5-1.5B-Instruct & 0.66 (+0.14) & 0.67 (+0.16) & 0.58 (+0.06) & 0.64 (+0.12) \\
    Qwen2.5-Coder-7B-Instruct & \underline{0.68} (+0.06) & 0.74 (+0.21) & \underline{0.64} (+0.16) & \underline{0.68} (+0.14) \\
    Qwen2.5-Coder-14B-Instruct & 0.64 (+0.02) & 0.65 (+0.06) & 0.57 (+0.08) & 0.62 (+0.06) \\
    Codestral-22B-v0.1 & 0.67 (+0.11) & 0.67 (+0.16) & 0.57 (+0.06) & 0.63 (+0.11) \\
    \midrule 
        \multicolumn{5}{c}{\textbf{Reverse binary targets}} \\
    \midrule
    \textit{Zero-shot} \\
    Qwen2.5-1.5B-Instruct & 0.54 & 0.49 & 0.46 & 0.49 \\
    Qwen2.5-Coder-7B-Instruct & 0.63 & 0.54 & 0.55 & 0.57 \\
    Qwen2.5-Coder-14B-Instruct & 0.65 & 0.59 & 0.57 & 0.60 \\
    Codestral-22B-v0.1 & 0.62 & 0.53 & 0.47 & 0.54 \\
    \midrule
        \textit{Fine-tuning} \\
    Qwen2.5-1.5B-Instruct & \underline{0.68} (+0.14) & 0.67 (+0.18) & 0.58 (+0.12) & 0.64 (+0.15) \\
    Qwen2.5-Coder-7B-Instruct & \textbf{0.69} (+0.06) & \underline{0.75} (+0.21) & \textbf{0.65} (+0.10) & \textbf{0.69} (+0.12) \\
    Qwen2.5-Coder-14B-Instruct & 0.64 (-0.01) & 0.65 (+0.06) & 0.53 (-0.04) & 0.60 (+0.0) \\
    Codestral-22B-v0.1 & 0.65 (+0.03) & 0.61 (+0.08) & 0.54 (+0.07) & 0.60 (+0.06) \\
    \bottomrule                        
  \end{tabular}
\end{table*}

\begin{table*}[htbp]
  \centering
  \small
  \caption{Results for Holdout dataset} 
  \label{tab:bench_results}           
  \begin{tabular}{@{} l c c c | c c c | c c c @{}}
    \toprule
    % Header row with multirow for Model
    \multirow{2}{*}{\textbf{Model}} & 
      \multicolumn{3}{c}{\textbf{Go}} & 
      \multicolumn{3}{c}{\textbf{Java}} & 
      \multicolumn{3}{c}{\textbf{Python}} \\
    \cmidrule(r){2-4} \cmidrule(r){5-7} \cmidrule(r){8-10}
    & \textbf{Cor.} & \textbf{TestCov} & \textbf{MutCov} &
      \textbf{Cor.} & \textbf{TestCov} & \textbf{MutCov} &
      \textbf{Cor.} & \textbf{TestCov} & \textbf{MutCov} \\
    \midrule                              
    \multicolumn{10}{c}{\textbf{Binary targets}} \\
    \midrule
    Qwen2.5-Coder-7B-Instruct (SFT) & 0.57 & 0.54 & 0.58 & 0.55 & 0.60 & 0.71 & 0.55 & 0.56 & 0.59 \\
    Codestral-22B-v0.1 (LoRa)       & 0.59 & 0.50 & 0.52 & 0.51 & 0.46 & 0.37 & 0.61 & 0.52 & 0.49 \\
    \midrule 
    \multicolumn{10}{c}{\textbf{Float targets}} \\
    \midrule
    Qwen2.5-Coder-7B-Instruct (SFT) & 0.62 & 0.57 & 0.61 & 0.61 & 0.61 & 0.68 & 0.64 & 0.61 & 0.67 \\
    Codestral-22B-v0.1 (LoRa)       & 0.61 & 0.54 & 0.58 & 0.53 & 0.46 & 0.50 & 0.56 & 0.53 & 0.55 \\
    \bottomrule                         
  \end{tabular}
\end{table*}

\begin{table*}[htbp]
  \centering
  \small
  \caption{Compare RM -RF with Execution-based metrics} 
  \label{tab:comparison_ef_soft}           

  \begin{tabular}{
      @{} l c c c | c c c | c c c | ccc @{}
  }
  \toprule
  \multirow{2}{*}{\textbf{Model}} & \multicolumn{3}{c}{\textbf{Go}} & \multicolumn{3}{c}{\textbf{Java}} & \multicolumn{3}{c}{\textbf{Python}} & \multicolumn{3}{c}{\textbf{Overall}} \\
  \cmidrule(r){2-4} \cmidrule(r){5-7} \cmidrule(r){8-10} \cmidrule(r){11-13}
     & \textbf{Exec.} & \textbf{RM -RF} & \textbf{$\Delta$} &
       \textbf{Exec.} & \textbf{RM -RF} & \textbf{$\Delta$} &
       \textbf{Exec.} & \textbf{RM -RF} & \textbf{$\Delta$}  &
       \textbf{Exec.} & \textbf{RM -RF} & \textbf{$\Delta$} \\
    \midrule                      
    GPT-OSS-120B & 0.56 & 0.40 & 0.16 & 0.02 & 0.18 & 0.16 & 0.20 & 0.24 & \underline{0.04} & 0.30 & 0.29 & \textbf{0.01} \\
    Devstral-small & 0.21 & 0.29 & \underline{0.08} & 0.01 & 0.05 & \textbf{0.04} & 0.05 & 0.07 & \textbf{0.02} & 0.10 & 0.16 & 0.06 \\
    Gemini-2.5-Flash & 0.17 & 0.41 & 0.24 & 0.01 & 0.14 & 0.13 & 0.04 & 0.22 & 0.18 & 0.08 & 0.27 & 0.19 \\
    Qwen3-Coder & 0.52 & 0.34 & 0.18 & 0.02 & 0.13 & 0.11 & 0.18 & 0.26 & 0.08 & 0.30 & 0.27 & \underline{0.03} \\   
    GPT-5 & 0.67 & 0.41 & 0.26 & 0.07 & 0.27 & 0.20 & 0.19 & 0.28 & 0.09 & 0.36 & 0.33 & \underline{0.03} \\ 
    DeepSeek-3.1 & 0.19 & 0.25 & \textbf{0.06} & 0.01 & 0.06 & \underline{0.05} & 0.05 & 0.19 & 0.14 & 0.09 & 0.17 & 0.08 \\ 
    \bottomrule                         
  \end{tabular}
\end{table*}

\begin{table}[htbp]
\small
\caption{Spearman's rank correlation coefficient}
\centering
\setlength{\tabcolsep}{4pt}
\renewcommand{\arraystretch}{1.05}
\begin{tabular}{lccc}
\toprule
        \textbf{Subset} & \textbf{NDCG} & \textbf{Coefficient} & \textbf{Description} \\
        \midrule
        Go & 0.79 & 0.4 & Moderate \\
        Java & 0.91 & 0.6 & Appreciable \\
        Python & 0.89 & 0.6 & Appreciable \\
        \midrule
        Overall & 0.86 & 0.74 & High \\

        \bottomrule
    \end{tabular}
    \label{tab:correlation}
\end{table}

\section{Evaluation}
\label{sec:eval}
Both fine-tuned models and zero-shot models were evaluated on the validation dataset, whose characteristics are in section: \ref{sec:dataset_structure}. The evaluation results and quality metrics are reported in Table \ref{tab:main_results}.

In addition, the testing was conducted on an extended sample; the methodology for this stage is described below.

The reward models produced predictions for two types of \emph{targets}: \emph{binary targets}, directly predicting correctness and whether coverage or mutability improved (\texttt{True}/\texttt{False}); and \emph{float targets}, where continuous gain values were binarized for evaluation (\texttt{True} if $> 0$, \texttt{False} otherwise).

The model performance was evaluated using a \emph{weighted F1-score}; class weights were computed based on label frequencies; each sample was assigned the weight of its true class; and weighted \emph{Precision} and \emph{Recall} were used to compute the final F1-score.

The best performing model, selected for highest mean accuracy, served as the \emph{reference model}. All other models were compared against it using the \emph{Wilcoxon signed-rank test}: for each example, the prediction differences were encoded as $+1$ (model correct, reference wrong), $-1$ (model wrong, reference correct), or $0$ (agreement). The test statistic $W$ was defined as the sum of ranks for cases where the model underperformed the reference. A value of $p$ $< 0.05$ indicated a statistically significant underperformance. The models were tested in the validation subset using this method. \texttt{Codestral-22B-v0.1} and \texttt{Qwen2.5-Coder-7B-Instruct} proved to be the best for binary and float targets.

The primary objective of \textbf{RM -RF} is to reduce latency while maintaining a comparable level of quality when generating and evaluating tests. To substantiate this claim, we conducted an empirical comparison between the evaluations produced by \textbf{RM -RF} and those derived from executing traditional testing methods and tools designed to measure both coverage and mutation kill rates. The fine-tuned \texttt{Qwen2.5-Coder-7B-Instruct} was evaluated as \textbf{RM -RF}. We computed the ratio of the useful generated test cases (it should be correct and increase either coverage or mutation kill rate). The findings are summarized in Table~\ref{tab:comparison_ef_soft}. Lower absolute differences $\Delta$ indicate that \textbf{RM -RF}' predictions closely align with actual execution outcomes, suggesting that \textbf{RM -RF} can effectively substitute full-scale test runs without losing accuracy. Furthermore, since \textbf{RM -RF} eliminates the need for building projects and executing source code, it offers substantial savings in terms of computational resources and processing time.

\section{Discussion}
Across settings, fine-tuning consistently improves predictive quality over zero-shot baselines (Table~\ref{tab:main_results}). The best average F1 is achieved by \texttt{Qwen2.5-Coder-7B-Instruct} after SFT (0.69) on binary targets, with a particularly strong gain on \texttt{\(\Delta\)TestCov} (0.76).

\emph{Binary} targets provide a robust and easy-to-learn supervision signal, yielding the strongest average results on the validation subset. \emph{Float} targets — binarized for evaluation — show slightly lower mean scores on validation but generalize better on the Holdout dataset (Table~\ref{tab:bench_results}), achieving equal or higher results for each metric. This suggests that continuous supervision helps the model capture graded relationships between test modifications and execution outcomes, improving transfer to unseen projects. Early experiments with \emph{reversed binary} targets exhibited behavior similar to the standard binary formulation but added complexity without consistent benefit; we therefore retain binary and float variants as the main targets. On the Holdout dataset, \textbf{RM -RF} also demonstrates consistent cross-language performance, maintaining stable accuracy across Go, Java, and Python. While \texttt{\(\Delta\)MutCov} prediction improvements remain most pronounced for Java (up to 0.71 on binary targets), \texttt{\(\Delta\)TestCov} prediction shows smaller variance across languages. These trends likely stem from structural and tooling differences—such as language-specific mutation operators, test frameworks, and code base organization. Together with the stronger holdout performance of float targets, these results indicate that the model captures generalizable relationships between code and test dynamics rather than overfitting to a single language or dataset distribution.

Full SFT on a 7B model outperforms PEFT LoRA on larger backbones in our regime (e.g., 7B SFT average F1 of 0.69 vs. 14B LoRA at 0.63 on binary targets). This points to a data/compute “sweet spot,” where full adaptation of a moderately sized model yields better alignment to our task than partial adaptation of larger models. LoRA remains competitive for some metrics and is attractive when memory budgets are tight, but in our experiments, SFT provided the most reliable gains.

Analyzing the model's performance on various subsets indicates that the subset \textbf{\textit{generated\_not\_correct}} (samples intentionally designed to include errors) poses the greatest challenge for determining test correctness. This difficulty might stem from the fact that the generated code has never been seen by any trained models before. A closer examination of error types reveals that issues like "Missed Dependencies" and "Duplicated Entity" are particularly hard for the model to detect, whereas errors such as "Invalid Constructor," "Undefined Entity," "Runtime Errors," and "Invalid Call" are relatively easier for the model to identify.

To further quantify alignment with execution-based evaluations, we computed both NDCG and Spearman’s rank correlation \cite{spearman1904reprinted} between \textbf{RM -RF} predictions and actual execution-derived metrics (Table~\ref{tab:correlation}). The results show moderate to high monotonic correlation across languages, with coefficients of 0.4 for Go and 0.6 for both Java and Python, and an overall correlation of 0.74. This indicates that \textbf{RM -RF} preserves relative ranking among test cases by quality, reinforcing its potential as a fast surrogate for execution-based evaluation.

Because \textbf{RM -RF} eliminates build and execution in the evaluation loop, it substantially reduces latency and compute cost while maintaining close alignment with execution-derived outcomes. This efficiency enables larger candidate pools for filtering, faster iterative refinement, and more frequent feedback in RL-based code optimization pipelines—particularly valuable when mutation testing is the primary bottleneck. To demonstrate this point, the 22B model predicts all samples within less than three hours, whereas complete building and executing steps take several days to yield the final results.

Building on these results, several extensions appear especially promising for future work:  
(1) scaling supervised fine-tuning (SFT) to larger models beyond 7B parameters;  
(2) integrating \textbf{RM -RF} directly into reinforcement learning loops as a reward signal for code generation;  
(3) exploring curriculum-style supervision that combines binary and float targets (e.g., starting with binary and subsequently fine-tuning on float deltas) to further stabilize mutation and coverage predictions;  
(4) comparing per-language fine-tuning of specialist reward models against a single multilingual variant.  
Together, these directions could further enhance both the scalability and precision of reward modeling for automated test generation.
\section{Limitations}

The present work has several limitations.

First, the dataset and models are limited by only three programming languages. However, the coverage across languages should be increased.

Second, we did not conduct the experiments in which one model is trained only for one programming language. From one point of view, it could increase the performance. From the other point of view, it would be more difficult to maintain such systems in practice.

Third, the reward model was not tested in the RL pipeline. It should be incorporated into the RL loop to increase the quality of the generated tests and speed up the training and generation process. Nevertheless, our experiments showed that the assessment from the \textbf{RM -RF} is close to the traditional tools that proves our concept.

Fourth, our dataset uses publicly available open-source code, potentially introducing partial data leakage if models encountered this code during pre-training. Nevertheless, we argue that the models had not been specifically trained to address these exact tasks; thus, even if they met similar code, it would not constitute data leakage.

\section{Conclusion}

In this work, we have introduced \textbf{RM -RF}, a novel lightweight reward model tailored for efficient evaluation of automatically generated unit tests without requiring compilation or execution. By predicting key metrics such as successful compilation, increased code coverage, and improved mutation kill rates directly from source and test code, \textbf{RM -RF} achieves significant reductions in latency and resource consumption. Our extensive experiments on a diverse multilingual dataset demonstrate its effectiveness across different programming languages (Java, Python, Go) and model configurations, yielding strong predictive performance with an average F1 score of 0.69. These results highlight \textbf{RM -RF}'s potential to serve as a practical alternative to conventional compile-and-run approaches, providing rapid, scalable, and accurate feedback essential for advancing automated test generation and reinforcement learning-driven code optimization efforts. 

\section{Data Availability}

All code, data, and experimental details necessary for reproducing our results are publicly available at \texttt{https://github.com/trndcenter/RM-RF-unit-tests}.

\bibliographystyle{IEEEtran}
\bibliography{saner-2026_references}

\end{document}